\documentclass[pdftex,twocolumn,epjc3]{svjour3}
\smartqed  % flush right qed marks, e.g. at end of proof

\usepackage[utf8]{inputenc}   
\usepackage[T1]{fontenc}

\usepackage{graphicx}
\usepackage{amsmath}
\usepackage{amsfonts}
\usepackage{amssymb}
\usepackage{mathrsfs}
\usepackage{longtable}
\usepackage{attrib}
\usepackage[rightcaption]{sidecap}

\usepackage{lineno}
\usepackage{color}

\sidecaptionvpos{figure}{t}
\usepackage{array}
\newcolumntype{P}[1]{>{\centering\arraybackslash}p{#1}}

\journalname{Eur. Phys. J. E}

\begin{document}

\title{Thickness of epithelia on wavy substrates: measurements and continuous models}

\author{Nicolas Harmand\thanksref{addr1} 
        \and
        Julien Dervaux\thanksref{e1,addr1}
        \and
				Christophe Poulard\thanksref{addr2} 
        \and
        Sylvie Hénon\thanksref{e2,addr1}
}

\thankstext{e1}{e-mail: julien.dervaux@u-paris.fr}
\thankstext{e2}{e-mail: sylvie.henon@u-paris.fr}

\institute{Universit\'e de Paris, CNRS, Mati\`ere et Syst\`emes Complexes, UMR 7057, F-75006 Paris, France\label{addr1}
           \and
           Laboratoire de physique des solides, CNRS, Université Paris-Saclay, UMR 8502, Orsay, France,\label{addr2}
}

\date{Received: date / Accepted: date}
% The correct dates will be entered by the editor

\maketitle

\begin{abstract}We measured the thickness of MDCK epithelia grown on substrates with a sinusoidal profile. We show that while at long wavelength the profile of the epithelium follows that of the substrate, at short wavelengths cells are thicker in valleys than on ridges. This is reminiscent of the so-called « healing length » in the case of a thin liquid film wetting a rough solid substrate. We explore the ability of continuum mechanics models to account for these observations. Modeling the epithelium as a thin liquid film, with surface tension, does not fully account for the measurements. Neither does modeling the epithelium as a thin incompressible elastic film. On the contrary, the addition of an apical active stress gives satisfactory agreement with measurements, with one fitting parameter, the ratio between the active stress and the elastic modulus.
\end{abstract}

\section{Introduction}
\label{intro}
In living organisms, epithelial tissues often display three-dimensional microstructures associated with curvature. Intestinal crypts and villi, pulmonary alveoli, renal tubules are some examples. Morphogenetic events are also associated with changes in the curvature of epithelial tissues, the archetypal example of which are embryo gastrulation and neurulation. \textit{In vitro} experiments on epithelia have long focused on flat 2D cultures. In recent years, strategies have been developed to perform experiments on curved epithelia \cite{salomon_contractile_2017,maechler_curvature_2018,fouchard_curling_2020,Luciano_2020,Harmand2021}. However there are still many open questions regarding the coupling between curvature and tissue morphology, identity and function. There is therefore an urgent need to develop quantitative studies of the shape of curved epithelia and models that account for the measurements. 
In this study we cultured MDCK epithelia on substrates with a sinusoidal profile, which allow to perform measurements on systems with controlled curvature and to have both positive and negative curvatures on the same sample. In order to vary the experimental conditions, we used sinusoidal substrates with different wavelength and amplitude values. In order to modulate the properties of the cells, we used wild-type MDCK cells and modified cells, which over-express the intercellular adhesion protein E-cadherin, coupled to the Green Fluorescent Protein (GFP). We evidence a thickening of the epithelia in the valleys of the substrate and a thinning on the ridges, for short wavelengths. To account for the measurements, we propose different models in which the epithelium is modeled as a continuous medium, either liquid or elastic. Surface tension and elasticity both oppose the deformation of the epithelium from a film with uniform thickness and tend to flatten the epithelium surface as compared to the substrate profile, but this is not sufficient to account for the experimental data. The addition of an active surface stress allows a satisfactory agreement between model and measurements.

\section{Epithelial thickness measurements on corrugated substrates}
\label{sec:1}
\subsection{Production of substrates with a sinusoidal profile}
\label{sec:substrates}
We created substrates with quasi-sinusoidal profiles by exploiting an elastic instability \cite{brau_multiple-length-scale_2011,Harmand2021}. A PDMS sample (8 cm $\times$ 3 cm $\times$ 2 mm, Sylgard 184 + 10\% cross-linker, DowCorning, cured overnight at 65$^{\circ}C$) was submitted to a uni-axial stretch in a home-designed device. Its surface was then oxidized by exposure to ozone generated by a UV lamp (Heraeus NNQ 20/18 U, 22W, distance 1cm from the sample), creating a rigid surface layer. 
When the strain was released, wrinkles formed because of the disparity between the Young's moduli of the oxidized layer and the bulk. The wavelength $\lambda$ and amplitude $A$ of the sinusoidal profile were controlled by varying the applied strain $\epsilon$ and the thickness of the oxidized layer, which was varied by changing the duration of UV exposure. Table \ref{wavy_moulds} shows the characteristics of the different moulds used in this study.

\begin{table}
% table caption is above the table
\caption{Characteristics of the templates}
\label{wavy_moulds}       % Give a unique label
% For LaTeX tables use
\begin{tabular}{llll}
\hline\noalign{\smallskip}
UV exposure & $\epsilon$ & $\lambda (\mu m)$ & A ($\mu m$)  \\
\noalign{\smallskip}\hline\noalign{\smallskip}
10 min & 25 \% & 40 & 5.6 \\
20 min & 20 \% & 54 & 6.7 \\
20 min & 30 \% & 49 & 10 \\
30 min & 20 \% & 110 & 11 \\
30 min & 30 \% & 96 & 16 \\
\noalign{\smallskip}\hline
\end{tabular}
\end{table}

These templates were used as moulds to obtain sinusoidal PDMS substrates. They were first silanized with vapor phase 1,1,2,2,-tridecafluoro-1,1,2,-tetrahydrooctyl-1-trichlorosilane to facilitate the release of the elastomer after curing. PDMS (Sylgard 184 + 10\% cross-linker, DowCorning) was then poured over the template, spin-coated for 90s at 450rpm to a thickness of $\sim 200 \mu m$, cured overnight at 65$^{\circ}C$, and finally peeled off.

\subsection{Cell culture}
\label{sec:cell_culture}
Madin-Darby canine kidney II (MDCK-II) cells from \textit{European Collection of Authenticated Cell Cultures} and MDCK-II cells genetically modified to stably express E-cadherin-GFP \cite{adams_mechanisms_1998} were cultured in \textit{Dulbecco’s Modifed Eagle Medium} (DMEM), supplemented with 10\% fetal bovine serum (FBS), 100U/mL \textit{penicillin-streptomycin}, and 50 $\mu$g/mL \textit{G418} (Geneticin) for cells expressing E-cadherin-GFP. In the following these cell lines are denoted MDCK-WT and MDCK-EcadGFP respectively.
For all experiments, flat or sinusoidal PDMS substrates were coated with fibronectin, 5 $\mu$g/mL in DMEM for 30 min at 37$^{\circ}$C. Confluent epithelia were then detached from culture flasks with trypsin-EDTA and cells were seeded onto the substrates at a density of 1900 cells/$mm^2$, which allowed confluence to be reached within 48h. 
The culture medium was renewed 48h after seeding.

\subsection{Immunofluorescence}
96h after seeding, cells were fixed with 4\% Paraformaldehyde (PFA) in phosphate-buffered saline (PBS).
The apical surface of the cells was labeled with mouse anti-gp135 (DSHB, reference 3F2/D8) \cite{herzlinger_mdck_1982} 1/25 for 20 min at room temperature and goat anti-mouse Dylight 549 (Abcam) for 30 min at room temperature; the nuclei were labeled with DAPI, 0.6 $\mu M$ in PBS for 20 min at room temperature;  F-actin was labeled with \textit{SiR-actin} (Spirochrome), 0.1 $\mu M$ in PBS for 12h at 4$^{\circ}$C.

In order to stain the substrates, 1/3 of the fibronectin was labeled with DyLight fast conjugation kit (Abcam). DyLight 488 was used for MDCK-WT and DyLight 549 for MDCK-EcadGFP.

\subsection{Image acquisition and processing}
\label{sec:image}
The samples were placed upside down under the microscope to avoid imaging through PDMS. Image acquisition was performed with Metamorph on a motorized inverted microscope (Leica DMI8) equipped with a 63X water immersion objective, a wide field spinning disk head CSU-W1 (Yokogawa - Andor), and a sCMOS Orca-Flash 4 V2+ camera (Hamamatsu), resulting in a field of view of $230 \mu m \times 230 \mu m$.
The sampling in the direction of the optical axis was $0.25 \mu$m.

MATLAB, ImageJ and MIJ \cite{sage_mij_2012} were used for image analysis. 

Since the geometry of the sinusoidal substrate makes 3D image analysis difficult, semi-automatic cell height measurements were limited to the ridges and the valleys of the substrate.
The program automatically created cross-sectional ($Z$) views along lines drawn by the user along each valley and ridge on one of the $XY$ images of the $Z$-stack.
Next, the gp135, F-actin and fibronectin channels were added, after renormalization of the intensities, to obtain cross-sectional images that clearly showed the apical and basal surfaces of the cells.
The thickness of the cells were finally automatically measured on the cross-sectional images. The values of the wavelength $\lambda$ and amplitude $A$ of the sinusoidal profile were not perfectly uniform within a given sample. They were measured by hand on each stack of images from the positions of ridges and valleys.
For flat substrates ($1/\lambda=0$), arbitrary parallel valleys and ridges were drawn.
Each measurement displayed in the following corresponds to mean values in a given stack of images. 

\begin{figure}
\centering
	\includegraphics[width=0.8\columnwidth]{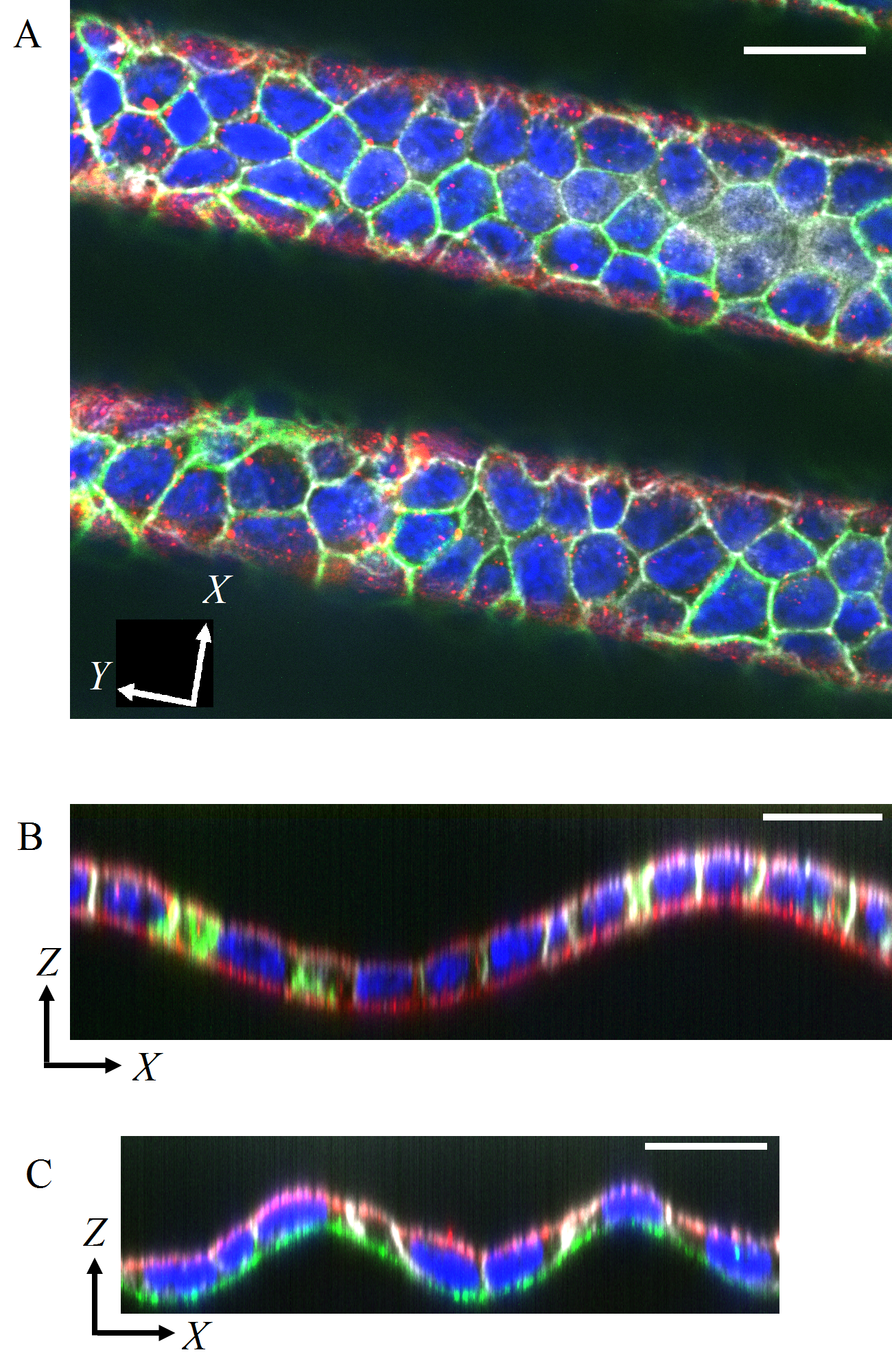}
\caption{MDCK epithelia grown on PDMS substrates with sinusoidal profiles. The nuclei appear in blue, F-actin in grey and the apical membranes in red. (A) Top view and (B) cross-section of MDCK-EcadGFP cells, E-cadherin appears in green and both apical membranes and fibronectin appear in red. (C) Cross-section view of MDCK-WT cells, fibronectin appears in green. Scale bars = 20 $\mu m$.}
\label{images_exp}
\end{figure}

\subsection{Epithelia are thicker in valleys than on ridges}
\label{sec:measurements}
\begin{figure}
\centering
	\includegraphics[width=0.8\columnwidth]{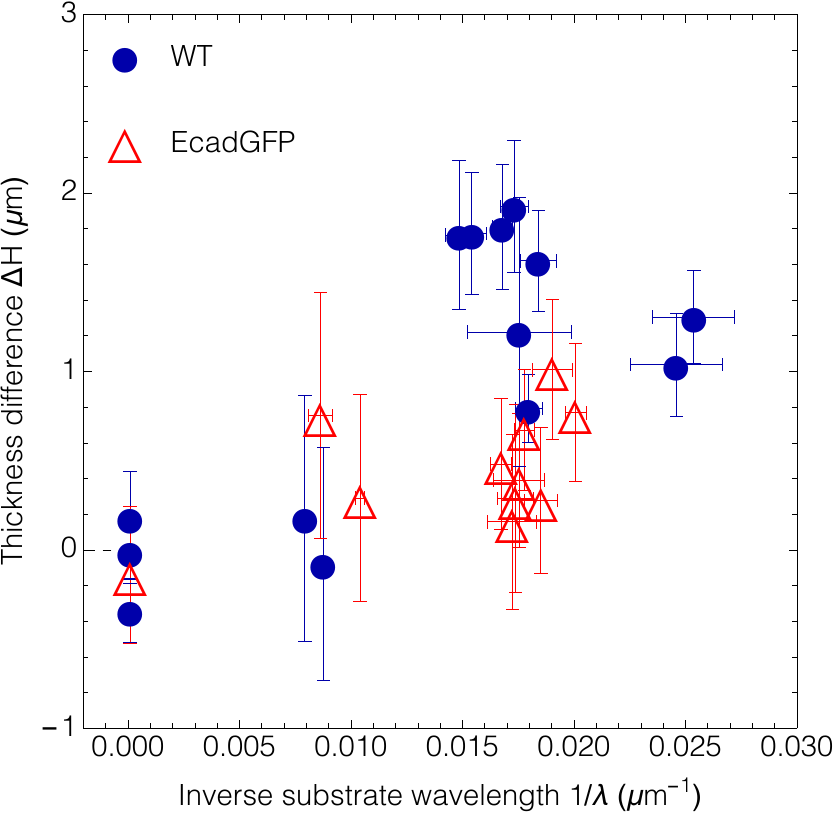}
\caption{Measured difference in the thickness of the epithelial cells in the valleys and on the ridges of the sinusoidal substrates as a function of the inverse of its wavelength. Error bars show the 95\% confidence intervals.}
\label{delta_H_vs_inv_lambda}      
\end{figure}

On substrates with a long wavelength $\lambda$, the shape of the epithelium closely followed that of the substrate, as appears in Fig. \ref{images_exp}.B. On the contrary, on substrates with a short wavelength $\lambda$, the cells were thicker in the valleys than on the ridges of the wavy substrates (Fig. \ref{images_exp}.C).
The measured thickness differences $\Delta H=H_{valley}-H_{ridge}$ are displayed as a function of $1/\lambda$ in Fig. \ref{delta_H_vs_inv_lambda}, for the two cell types, MDCK-WT and MDCK-EcadGFP. For flat substrates, $\Delta H$ was measured as a control for parallel arbitrary valleys and ridges: $\lvert \Delta H \rvert$ is smaller than 0.35 $\mu$m, which is therefore the sensitivity of the measurement. For long wavelengths (typically $\lambda>100\mu m$), $\Delta H \simeq 0$. On the contrary, for short wavelengths $\Delta H > 0$. Note that $\Delta H$ is smaller for MDCK-EcadGFP than for MDCK-WT, while the average epithelium is a little larger for MDCK-EcadGFP than for MDCK-WT (typically 5 \textit{vs} 6 $\mu m$).

\section{Models}
\label{sec:2}
\subsection{Liquid film with interfacial tensions}
\label{sec:2.1}

\paragraph{Thin wetting liquid film on a rough substrate.}
Our observation is reminiscent of the properties of a thin liquid film completely wetting a rough solid surface \cite{deGennes2003,Andelman1988}, with an optimal value $H_0$ of the film thickness on a flat substrate, governed by long range attractive forces such as van der Waals interactions. When the liquid is on a rough substrate, long wavelength undulations of the liquid surface follow that of the solid surface, while short wavelength undulations are damped, in order to reduce the energy cost of creating liquid surface. More precisely the local equilibrium thickness $H$ of the thin film is determined by the balance between two competing energies per unit area: the surface tension of the liquid $\gamma$ and an energy depending on the film thickness, $P(H)$, related for instance to van der Waals interactions \cite{deGennes2003,Andelman1988}. A healing length can then be defined:

\begin{equation}
\xi_e=\left(\frac{1}{\gamma}\left(\frac{d^2 P}{dH^2}\right)_{H=H_0}\right)^{-1/2}
\label{xi_e}
\end{equation}

If the substrate is rough, with a typical wavelength $\lambda$, the surface of the liquid film follows the solid roughness if $\lambda \gg \xi_e$ and is smooth if $\lambda \ll \xi_e$. Furthermore when the modulation of the substrate is purely sinusoidal with wavelength $\lambda$ and amplitude $A$, the modulation of the liquid surface is expected to vary sinusoidally, with an amplitude (see Appendix A for details):
\begin{equation}
a=\frac{A}{1+\left(\frac{2 \pi \xi_e}{\lambda} \right)^2}
\label{a_over_A_vs_lambda}
\end{equation}

\paragraph{Modeling an epithelium as a thin liquid film.}
\label{sec:2.2}

Following previous works \cite{honda_three-dimensional_2004,Hannezo2014,Misra2016,Harmand2021}, we model the confluent epithelium as a cohesive sheet of identical cells and we assume that the state of the cell is determined by the minimization of an energy containing only surface tension terms:

\begin{equation}
E_{cell} = \gamma_{cs} S_{cs} + \gamma_{cm} S_{cm} + \frac{1}{2} \gamma_{cc} S_{cc}
\end{equation}

\noindent where $S_{cs}$, $S_{cm}$ and $S_{cc}$ are the areas of the interfaces between the cell and the substrate, the medium and the neighboring cells respectively, and where the $\gamma$s are the corresponding energies per unit area. For prismatic cells with basal area $S$, thickness $H$, volume $SH=V$, and defining the shape index $\alpha$ as the ratio between the cell perimeter and the square root of the cell area, $\sqrt{S}$, $E_{cell}$ is rewritten:
\begin{equation}
E_{cell} = S \left( \gamma_{cs} + \gamma_{cm} +  \frac{\alpha \gamma_{cc}}{2 \sqrt{V}} H^{3/2} \right)
\label{E_cell}
\end{equation}

With the additional assumption that the volume $V$ of the cells is a constant, the optimal value of $H$ is the one for which $E_{cell}(H)$ is minimal:
\begin{equation}
H_0 = V^{1/3} \left( \frac{4}{\alpha}\frac{\gamma_{cs} + \gamma_{cm}}{\gamma_{cc}} \right)^{2/3}
\end{equation}

The energy per unit area of the epithelium $E_{cell}/S$ (\textit{cf.} Eq. \ref{E_cell}) is similar to that of a thin liquid film with:
\begin{equation}
P(H) = \frac{\alpha \gamma_{cc}}{2 \sqrt{V}} H^{3/2}
\end{equation}

When grown on a substrate with a sinusoidal profile, an epithelium is therefore expected to follow the behavior described by Eq. \ref{a_over_A_vs_lambda}, with a healing length equal to:
$$\xi_e = V^{1/3} \left(\frac{2\gamma_{cm}}{3\gamma_{cc}}\right)^{1/2} \left(\frac{4}{\alpha}\right)^{2/3} \left(\frac{\gamma_{cs} + \gamma_{cm}}{\gamma_{cc}} \right)^{1/6}~~~~~~~~~~$$
\begin{equation}
~~~= H_0 \left(\frac{2}{3}\frac{\gamma_{cm}}{\gamma_{cs} + \gamma_{cm}}\right)^{1/2}
\label{xi_e-cell}
\end{equation}

If we drop the assumption that the volume of the cell constant, then a compressibility term, $B(V-V_0)^2$, has to be added in Eq. \ref{E_cell}. The conclusions remain qualitatively the same, except for the expressions of $P(H)$ and $\xi_e$, which contain a term depending on $B$.

\paragraph{Comparison with measurements.} In experiments on MDCK cells, we measured the difference $\Delta H$ in the thickness of the epithelium between the valleys and the ridges of substrates with sinusoidal profile (see sect. \ref{sec:measurements}). Following Eq. \ref{a_over_A_vs_lambda}, it is expected to vary as (see Appendix A for details):
\begin{equation}
\Delta H =\frac{A}{1+\left(\frac{\lambda}{2 \pi \xi_e} \right)^2}
\label{eq_delta_H_vs_lambda}
\end{equation}

\begin{figure}
\centering
	\includegraphics[width=0.8\columnwidth]{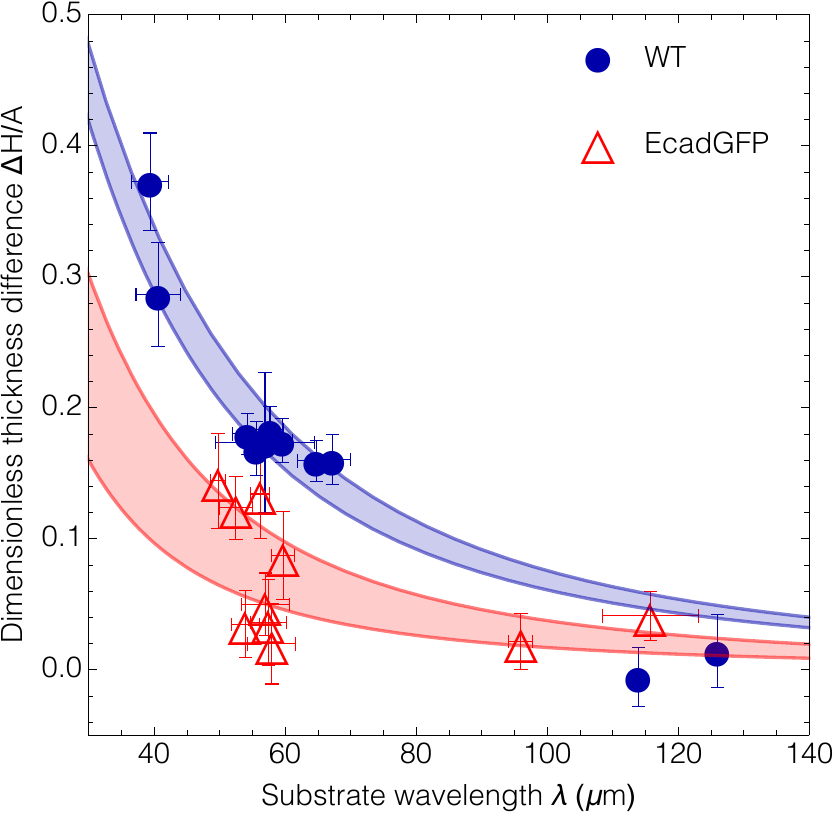}
\caption{Experimental values of the difference $\Delta H$ in the thickness of the epithelial cells in the valleys and on the ridges of the substrates, divided by the amplitude $A$ of the sinusoidal profile, and their best fits with Eq. \ref{eq_delta_H_vs_lambda}. Error bars show the 95\% confidence intervals.}
\label{delta_H_over_A_vs_lambda}      
\end{figure}

Fig. \ref{delta_H_over_A_vs_lambda} displays the measured values of $\Delta H / A$ as a function of $\lambda$ for the two cell lines and their best fits with Eq. \ref{eq_delta_H_vs_lambda}. The model is in fair agreement with the measurements. The value of the fitting parameter $\xi_e$ is equal to 4.3 $\pm$ 0.3 $\mu$m for MDCK-WT cells and to 2.6 $\pm$ 0.6 $\mu$m for MDCK-EcadGFP cells. Using Eq. \ref{xi_e-cell} and using values of $H_0$ measured as the mean value of the epithelia thicknesses, namely 5.1 $\mu m$ for MDCK-WT epithelia and 6.1 $\mu m$ for MDCK-EcadGFP epithelia, this gives $\gamma_{cs}/\gamma_{cm} \simeq$ -0.06 for MDCK-WT and 2.5 for MDCK-EcadGFP cells. The apical surface tension $\gamma_{cm}$ is mainly generated by the tension of the acto-myosin cortex of the apical face \cite{salbreux_actin_2012,chugh_actin_2018} while the cell-substrate surface tension $\gamma_{cs}$ has mainly two contributions, a positive cortex tension and a negative cell-substrate adhesion energy; $\gamma_{cs}$ is thus expected to be smaller than $\gamma_{cm}$. The value $\gamma_{cs}/\gamma_{cm} \simeq$ -0.06 is therefore consistent with the physical meaning of the parameters, with the adhesion energy of the cells to the substrate a little higher than the tension of the basal cortex; however $\gamma_{cs}/\gamma_{cm} \simeq$ 2.5 is not. In conclusion, although the measured behavior of the thickness of MDCK epithelia on sinusoidal substrates is similar to that of thin liquid wetting films, and in qualitative agreement with a surface tension only model, the model cannot account for all the measurements.

\subsection{Thin elastic film}
\label{sec:elastic}

We now examine the role of elasticity. The epithelium is modeled as a continuous, incompressible soft elastic layer, with shear elastic modulus $\mu$, resting over an infinitely stiff corrugated substrate. In the reference configuration, we assume that the layer is flat, with thickness $H_0$, and stress-free. The layer in the reference configuration is thus described by the region $-\infty < X <\infty$ and $0 < Z < h_1(X)=H_0$. 

For simplicity, we denote $\gamma$ the energy per unit area associated with apical surface of the epithelium, that was denoted $\gamma_{cm}$ in the previous sect.

The interface between the rigid substrate and the epithelium is now deformed from a flat plane into a sinusoidal profile:

\begin{equation}
h_0(X) = \frac{A}{2}  \cos\left( \frac{2 \pi X}{\lambda} \right)
\label{sinus_profile}
\end{equation}

Because of this imposed boundary condition, the layer will deform. However the elasticity and the surface tension of the apical surface both oppose this deformation. As a consequence, the apical surface will always remain "flatter" (less deformed) than the interface between the rigid substrate and the epithelium. 

In the deformed state a point of the epithelium with coordinates $\vec{X}=\{X,Z\}$ in the reference configuration is mapped to a point $\vec{x} = \vec{X}+\vec{u}(\vec{X})$ where $\vec{u}(\vec{X})=\{u_x(X,Z),u_z(X,Z)\}$ is the displacement field. Within the framework of linear elasticity theory, and under the incompressibility condition mentioned above:

\begin{equation}
\nabla \cdot \boldsymbol{\sigma}=\nabla \cdot \boldsymbol{\sigma}^{el}= \vec{0}\,\,\, \mbox{and} \,\,\, \vec{\nabla} \cdot \vec{u} =0
\label{Navier}
\end{equation}

The stress tensor $\boldsymbol{\sigma}^{el}$ is related to the strain tensor $\boldsymbol{\epsilon} = ( \nabla \vec{u} + ( \nabla \vec{u} )^T)/2$ via Hooke's law:

\begin{equation}
\boldsymbol{\sigma}^{el}= 2 \mu \boldsymbol{\epsilon} - p \bold{I}
\label{Hooke}
\end{equation}

\noindent where $\bold{I}$ is the identity matrix, $p(X,Z)$ is the pressure field and $\mu$ is the shear modulus of the epithelium. Using this constitutive equation, the equilibrium equation can also be written in terms of displacement field:

\begin{equation}
\mu \triangle\vec{u} - \vec{\nabla} p = 0\,\,\, \mbox{and} \,\,\, \vec{\nabla} \cdot \vec{u} =0
\label{displacement}
\end{equation}

The Navier equations must be supplemented with appropriate boundary conditions. At the bottom surface $Z=0$, the epithelial layer is bound to the corrugated surface and thus:

\begin{equation}
u_x (X,0)= 0 \,\,\, \mbox{and} \,\,\, u_z(X,0) = h_0(X)
\label{boundary}
\end{equation}

At the apical surface, the continuity of the total stress must be enforced:

\begin{equation}
\boldsymbol{\sigma}\cdot \vec{n} = \gamma \vec{n}\cdot (\vec{\nabla}\vec{n})
\label{normal_surface_stress}
\end{equation}

\noindent which implies the following balance in the tangential plane:

\begin{equation}
\mu \left(\frac{\partial u_x}{\partial Z} + \frac{\partial u_z}{\partial X}  \right) = 0 \,\,\, \mbox{at} \,\,\, Z = H_0
\label{tangent_surface_stress}
\end{equation}

\noindent while the normal stress balances the Laplace pressure at the free surface:

\begin{equation}
2\mu \frac{\partial u_z}{\partial Z} - p  = \gamma \frac{\partial^2 u_z}{\partial X^2} \,\,\, \mbox{at} \,\,\, Z= H_0
\label{Laplace}
\end{equation}

In order to make simple theoretical prediction on the shape of the epithelium, we shall make the simplifying assumption that the amplitude of the sinusoidal substrate is small. Skipping the details of the simple resolution procedure (presented in Appendix B), we find that the height of the deformed surface, denoted $\zeta(X)$, is given by the following expression:

\begin{equation}
\zeta(X) = H_0 + h_0(X) \,\, F\left(\frac{2\pi H_0}{\lambda},\frac{\pi \gamma }{\mu \lambda}\right)
\label{zeta00}
\end{equation}

\noindent where the function of two variables $F(x,y)$ is defined as

\begin{equation}
F(x,y) = \frac{2\left\{\cosh{\left(x\right)} + x \sinh{\left(x \right)} \right\}}{1 + 2 x^2\left(1 - \frac{y}{x} \right) + \cosh{\left(2 x\right)} + y \sinh{\left(2 x\right)}} 
\label{zeta0}
\end{equation}

The above formula involves two dimensionless parameters: $2\pi H_0 /\lambda$, which quantifies the influence of the geometry on the deformation, and $\pi \gamma/\mu \lambda$, a mechanical parameter which quantifies the relative strength of the surface energy and the bulk elastic energy. The function $F$ has values in the interval $[0,1]$ and the value $0$ (a flat free surface) is reached for both limits $2\pi H_0 /\lambda \rightarrow \infty$ and $\pi \gamma/\mu \lambda \rightarrow \infty$. These two limits correspond respectively to: (i) the thickness of the epithelial layer is much larger than the wavelength of the corrugated substrate, (ii) the surface energy of the epithelium (per unit of wavelength of the corrugated substrate) is much larger than the bulk elastic energy. The above formula also indicates that, in the limit of negligible surface energy, the shape of the epithelium follows that of the substrate, and is in fact only controlled by the geometry, independently of its elastic properties. From the previous result, the thickness $H(X)=\zeta(X)-h_0(X)$ is:

\begin{equation}
H(X)= H_0 + h_0(X) \, \left[F\left(\frac{2\pi H_0}{\lambda},\frac{\pi \gamma }{\mu \lambda}\right) -1 \right] 
\end{equation}

Experiments have focused on the thickness difference of the epithelium between valleys and ridges of the sinusoidal surface, which reduces to:

\begin{equation}
\frac{\Delta H}{A} = 1-F\left(\frac{2\pi H_0}{\lambda},\frac{\pi \gamma }{\mu \lambda}\right)
\label{elastic_model}
\end{equation}

\begin{figure}
\centering
	\includegraphics[width=0.8\columnwidth]{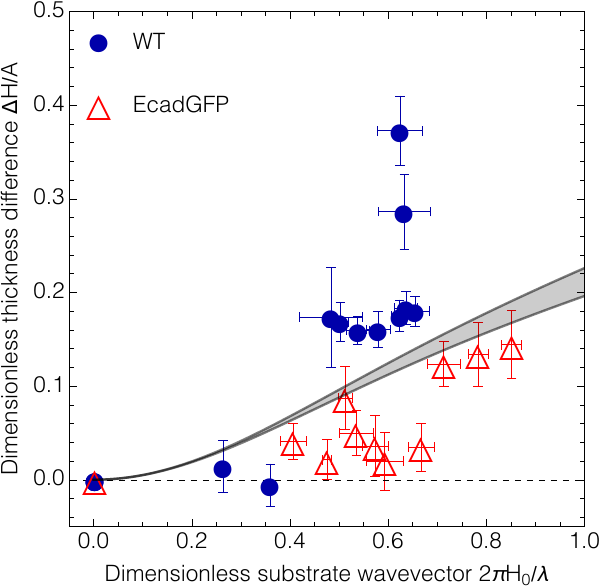}
\caption{Experimental values of the dimensionless thickness difference $\Delta H / A$, as a function of the dimensionless substrate wave vector $2\pi H_0/\lambda$, and plots of the model, $\left[1-F\left(\frac{2\pi H_0}{\lambda},\frac{\pi \gamma}{\mu \lambda}\right)\right]$ (\textit{cf.} Eq. \ref{elastic_model}). The gray area corresponds to the interval $\pi \gamma/\mu \lambda = 0-0.08 $ which is the largest interval that can be reached assuming the smallest value of $\lambda$ (40$\mu$m) and the highest bound for $\gamma/\mu$ of $1\mu$m.} Error bars show the 95\% confidence intervals.
\label{fig_elastic_model}   
\end{figure}

Fig. \ref{fig_elastic_model} displays measured values of $\Delta H / A$ as a function of $2\pi H_0/\lambda$ for MDCK-WT and MDCK-EcadGFP epithelia, along with plots of $\left[1-F\left(\frac{2\pi H_0}{\lambda},\frac{\pi \gamma }{\mu \lambda}\right)\right]$ with values of $\gamma /\mu$ between 0 and 5 $\mu$m. The best fit is obtained with $\gamma/\mu \sim18\mu$m for MDCK-WT cells, and would require a negative value of $\gamma/\mu$ for MDCK-EcadGFP cells. Since the surface tension $\gamma$ of the cell is smaller than 1 mN/m \cite{tinevez_role_2009,pietuch_mechanics_2013,cartagena-rivera_apical_2017} while the shear modulus of the epithelium is of the order of 1kPa \cite{balland_power_2006,nehls_stiffness_2019}, $\gamma/\mu$ is smaller than 1 $\mu$m. With such a value of $\gamma/\mu$, the model cannot account for the measurements.

In conclusion, modeling the epithelium by a thin elastic film with surface tension is in qualitative but not in quantitative agreement with our measurements, given the experimental values of shear modulus and apical tension.

\subsection{Thin elastic film with apical active stress}
\label{sec:active}

In addition to the model proposed in sect. \ref{sec:elastic}, we now assume that the epithelium is subjected to a 2D active stress field $\boldsymbol{\sigma}^{act}$ located in the vicinity of the apical surface and parallel to the surface. Indeed, cells are under self-created active stress, and in epithelia the internal active stress is more contractile at the apical surface, mainly due to a contractile circumferential actomyosin belt lining adherens junctions \cite{cartagena-rivera_apical_2017,lecuit_e-cadherin_2015,okuda_apical_2013,Balasubramaniam_2021,Harmand2021}. We postulate an additive decomposition of the total stress $\boldsymbol{\sigma}$ into elastic $\boldsymbol{\sigma}^{el}$ and active $\boldsymbol{\sigma}^{act}$ parts:

\begin{equation}
\boldsymbol{\sigma} = \boldsymbol{\sigma}^{el} + \boldsymbol{\sigma}^{act}
\end{equation}

In terms of components, our assumptions about active stress implies that $\sigma_{xz}^{act} =\sigma_{zx}^{act} = \sigma_{zz}^{act} = 0$ while $\sigma_{xx}^{act} $ does not vanish only in a thin layer of thickness $\delta$ below the free surface:

\begin{equation}
   \sigma_{xx}^{act} = 
\begin{cases}
   \sigma^{act} & \text{if } H_0-Z\leq \delta\\
    0,              & \text{otherwise}
\end{cases}
\end{equation}

Note that for a flat layer the active stress does not induce any deformation of the cell layer since it is divergence free. For a non-flat substrate however, the active stress, along with the elasticity of the epithelium and the apical surface tension, will oppose the deformation imposed by the boundary condition on the substrate.

We use the same notations as in sect. \ref{sec:elastic}. Note that the only non-zero component of the active stress is $\sigma_{xx}$ which is solely a function of $Z$ and thus the active stress is divergence-free ($\nabla \cdot \boldsymbol{\sigma}^{act} = \vec{0}$). Consequently the force balance (Eq. \ref{Navier}) remains unchanged. Furthermore, because of the peculiar form of the active stress, it does not enter the normal force balance at the free surface of the epithelium and Eq.\ref{Laplace}, together with Eq. \ref{boundary} also holds. However the balance of shear stress at the free surface is modified by the presence of the active stress and is replace by:

\begin{equation}
\mu \left(\frac{\partial u_x}{\partial Z} + \frac{\partial u_z}{\partial X}  \right) =\sigma^{act} \frac{\partial u_z}{\partial X} \,\,\, \mbox{at} \,\,\, Z= H_0
\label{shear}
\end{equation}

The resolution of the equations (shown in Appendix B) leads to:

\begin{equation}
\zeta(X) = H_0 + \frac{A}{2}  \cos\left( \frac{2 \pi X}{\lambda} \right) G\left(\frac{2\pi H_0}{\lambda},\frac{\pi \gamma }{\mu \lambda}, \frac{\sigma^{act}}{2\mu}\right)
\label{zeta1}
\end{equation}

\noindent where the function of three variables $G(x,y,z)$ is defined as

\begin{equation}
G(x,y,z) = \frac{2\left\{\cosh{\left(x\right)} + x \sinh{\left(x \right)} \right\}}{1 + 2 x^2\left(1-z - \frac{y}{x} \right) + \cosh{\left(2 x\right)} + y \sinh{\left(2 x\right)}} 
\label{zeta2}
\end{equation}

The new dimensionless parameter $\sigma^{act}/(2\mu)$ is a measure of the strength of the active apical stress of the epithelium, as compared to its elastic modulus. The function $G$ takes values between 0 and 1 and the value $0$ (flat apical surface) is reached for any of the limits: $2\pi H_0 /\lambda \rightarrow \infty$, $\pi \gamma/\mu \lambda \rightarrow \infty$ or $\sigma^{act}/(2\mu) \rightarrow \infty$. Furthermore, in the limit where the surface energy and active stress are negligible, the shape of the epithelium is only controlled by the geometry, independently of its elastic properties. From the previous result, we infer:

\begin{equation}
\frac{\Delta H}{A} = 1-G\left(\frac{2\pi H_0}{\lambda},\frac{\pi \gamma }{\mu \lambda},\frac{\sigma^{act}}{2\mu}\right)
\label{Delta_H_over_A_act}
\end{equation}

\begin{figure*}
  \includegraphics[width=\textwidth]{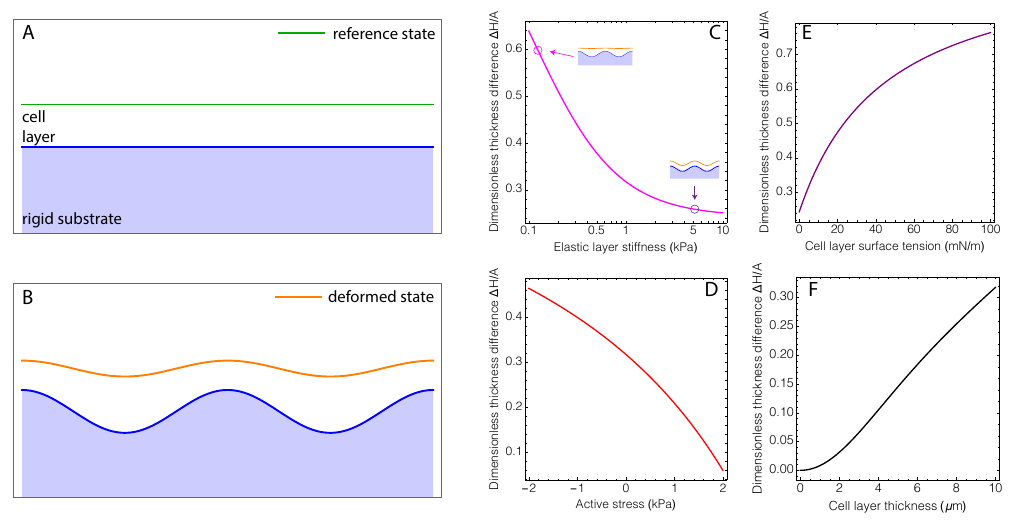}
\caption{Elastic model with active stress and surface tension for epithelia. (A) Reference state: a soft flat epithelium sits on top of a rigid substrate. (B) Final shape of the epithelium when the substrate is deformed into a sinusoidal profile. Panels (C) to (F): respective contribution of stiffness, active surface stress, apical tension and epithelium thickness to the profile of the epithelium on sinusoidal substrate. (C) Log-linear plot of the dimensionless thickness difference $\Delta H/A$ as a function of the epithelium stiffness $\mu$ in kPa. The thickness $H_0$ of the epithelium is set at 10 $\mu$m, its apical tension $\gamma$ at 5mN/m, the active stress $\sigma^{act}$ at 0 and the wavelength $\lambda$ of the sinusoidal substrate at 50 $\mu$m. The insets show the limits for a very soft and a very stiff epithelium. (D) Effect of $\sigma^{act}$ on $\Delta H/A$; $H_0$=10 $\mu$m, $\gamma$=5mN/m, $\mu$=1kPa, and $\lambda$=50 $\mu$m. (E) Effect of $\gamma$ (in mN/m) on $\Delta H/A$; $H_0$=10 $\mu$m, $\mu$=1kPa, $\sigma^{act}$=0 and $\lambda$=50 $\mu$m. (F) Effect of $H_0$ (in $\mu$m) on $\Delta H/A$;  $\gamma$=5 mN/m, $\mu$=1kPa, $\sigma^{act}$=0 and $\lambda$=50 $\mu$m.}
\label{illustr_model}
\end{figure*}

When the substrate is deformed into a sinusoidal profile, the final shape of the epithelium results from a balance between the epithelium surface tension, its elastic stiffness, the magnitude of the apical active stress and the geometry of the deformation. The respective effect of the geometry, the apical tension, the elastic stiffness of the epithelium and the magnitude of the apical active stress on the shape of the epithelium are illustrated in Fig. \ref{illustr_model}, panels C to F. Fig. \ref{illustr_model}.C illustrates the influence of the epithelium stiffness. When the epithelium is very soft ($\mu \rightarrow$ 0), the free surface tends to remain flat ($\Delta H/A \rightarrow$ 1), whereas a very stiff epithelium tends to closely follow the shape of the substrate/epithelium interface ($\Delta H/A \rightarrow$ 0) as shown in the insets. Fig. \ref{illustr_model}.D illustrates the influence of the active stress on the dimensionless thickness difference $\Delta H/A$. Compressive (negative) apical stress values flattens the surface of the epithelium while extensile (positive) apical stress leads to more pronounced oscillations of the free surface of the epithelium. Fig. \ref{illustr_model}.E illustrates the effect of the apical tension on the shape of the epithelium. Increasing the surface tension flattens the interface. However, because the surface tension of the epithelium typically lies within the range 0-5mN/m, the effect of surface tension on the shape of the epithelium is rather small. Finally, as illustrated on Fig. \ref{illustr_model}.F, increasing the thickness of the epithelium flattens its free surface.

\begin{figure}
\centering
	\includegraphics[width=0.8\columnwidth]{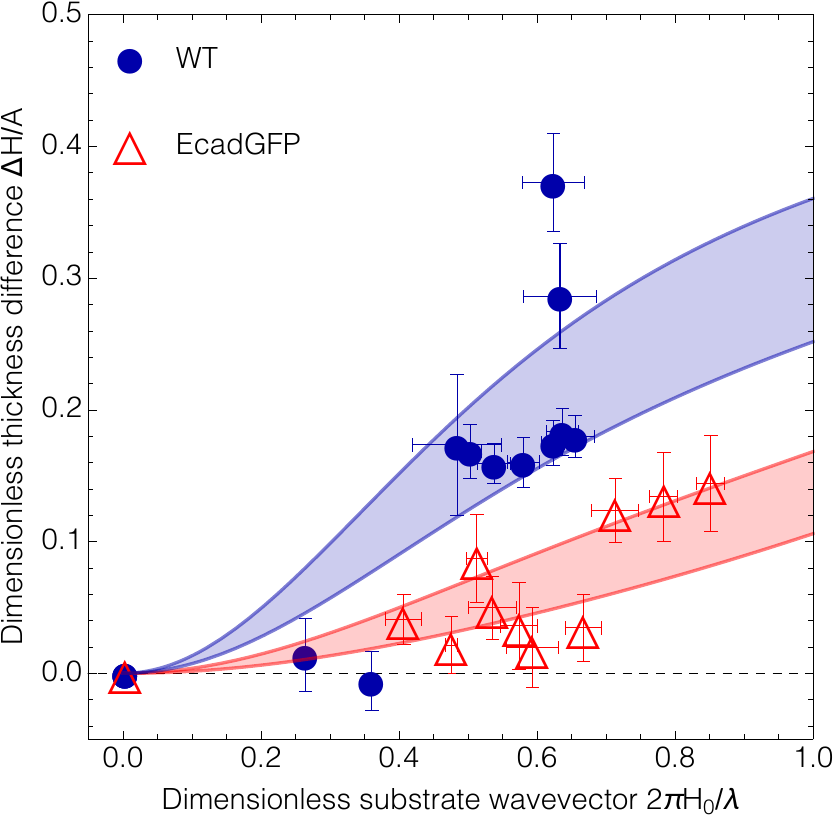}
\caption{Experimental values of the dimensionless thickness difference $\Delta H / A$, as a function of the dimensionless substrate wave vector $2\pi H_0/\lambda$, and best fits with $\left[1-G\left(\frac{2\pi H_0}{\lambda},0,\frac{\sigma^{act}}{2\mu}\right)\right]$ (\textit{cf.} Eq. \ref{Delta_H_over_A_act}), with $\sigma^{act}$ as a fitting parameter. Error bars show the 95\% confidence intervals.}
\label{fig_active_stress_model}
\end{figure}

In the experiments, as previously discussed, $\pi \gamma/\mu<$ 1 $\mu$m, hence $\pi \gamma/\mu \lambda<$ 0.08 for all values of $\lambda$ (40-130 $\mu$m); we shall therefore set $\pi \gamma/\mu \lambda = 0$. Fig. \ref{fig_active_stress_model} displays the measured values of $\Delta H/A$ as a function of $2 \pi H_0/\lambda$ for the two data sets, and their best fits with Eq. \ref{Delta_H_over_A_act}, with $\pi \gamma/\mu \lambda=0$ and $\sigma^{act}/(2\mu)$ as a fitting parameter:  $\sigma^{act}/(2\mu)$ = -0.56$\pm$0.31 for MDCK-WT and +0.23$\pm$0.11 for MDCK-EcadGFP. Since the apical actin belt is contractile, $\sigma^{act}$ is expected to be negative. For MDCK-WT epithelia, $\mu \sim$ 1 kPa and hence $\sigma^{act}$ is in the kPa range. With the hypothesis that $\sigma^{act}$ mainly originates from the contractile apical actin belt, which has a thickness $h$ of a fraction of $\mu$m, and acts on the periphery of the cell, with a perimeter $P \simeq$ 35 $\mu$m (for MDCK cells), this gives a line tension $\Lambda = \lvert \sigma^{act} \rvert h P \sim$ 1 nN. This is in the range of the values, 0.1-10 nN, that have been measured in \textit{Drosophila} embryos \cite{bambardekar_direct_2015,solon_pulsed_2009}. For MDCK-EcadGFP cells, the fit gives $\sigma^{act}$ > 0, which may seem inconsistent. Nevertheless individual cells are contractile but an epithelium, when modeled as a 2D system, can be either contractile or extensile \cite{Saw_2017} \cite{Blanch-Mercader_2018}. In our 3D description of the epithelium, $\sigma^{act}$ can be seen as the over-contractility of the apical side of the epithelium with respect to the average. A negative value of $\sigma^{act}$ thus corresponds to an epithelium less contractile at the basal part than at the apical part, and hence to an extensile epithelium, in accordance with measurements made with MDCK-WT epithelia \cite{Saw_2017}. In MDCK-EcadGFP cells, E-cadherins are expected to be over-expressed, since E-cadherin-GFP is expressed on top of endogeneous E-cadherin. One would thus expect the apical actin belt, which lines adherens junctions, to be reinforced in MDCK-EcadGFP cells as compared to MDCK-WT. Nevertheless, overexpression of E-cadherin is also expected to increase the cohesiveness of the epithelium, and hence to increase its global stiffness and contractility \cite{baronsky_reduction_2016,arslan_holding_2021}. The net effect on the value of $\sigma^{act}/\mu$ and $\pi \gamma/\mu$ is thus difficult to predict. Our measurements are in favor of a decrease in the value apical contractility with respect to the overall contractility, leading to $\sigma^{act}$>0. It would be interesting to test whether MDCK-EcadGFP epithelia are indeed contractile, like MDCK epithelia that do not express E-cadherin \cite{Balasubramaniam_2021}. 

\section{Discussion and conclusion}

In conclusion, we developed an experimental set-up that allowed us to grow epithelia on curved substrates of controlled shape, and with both negative and positive curvature close to each other in the same sample. We were able to demonstrate that the parts of the epithelium with positive curvature (in valleys of the substrate) were thicker than those with negative curvature (on the ridges). While a lot of models of epithelia focus on the properties of individual cells that constitute the tissue \cite{okuda_apical_2013,Hannezo2014,hannezo_interplay_2016,rupprecht_2017,rozman_collective_2020,Harmand2021}, we tested the ability of different continuous mechanics models to account for our measurements. We showed that apical surface tension and elasticity of the epithelium both qualitatively account for the measurements but not quantitatively. The addition of an active surface stress does. Building the correspondence between cellular and continuous models would be fruitful. Although the apical active stress probably originates from the so-called actin belt and the shear elastic modulus is somehow related to the cell-cell tension, the exact correspondence is not straightforward since the mechanical properties of a tissue arise not only from the properties of the cells, but also from their shapes and their relative movements when the tissue is deformed \cite{Murisic_2015,Tlili_2015}.

\begin{acknowledgements}
We acknowledge the ImagoSeine core facility of the Institut Jacques Monod (member of the France Bio\-Imaging, ANR-10-INBS-04) and the LabEx “Who Am I?” ANR-11-LABX-007 (Transition post-doctoral program). This work was funded in part by the French Agence Nationale de la Recherche Grant “AdhesiPS” ANR-17-CE08-0008.
\end{acknowledgements}

\section*{Author contribution statement}

\noindent NH performed all the experiments, image processing, data analysis, and calculations of sect. \ref{sec:2.1}. JD performed all the calculations and analysis of sects. \ref{sec:elastic} and \ref{sec:active}. CP designed and built the set-up for the production of corrugated substrates. SH designed and supervised the work and wrote the paper.

\section*{Appendix}
\subsection*{A. Healing length of thin liquid films}
Following \cite{deGennes2003,Andelman1988}, we describe the properties of a thin film wetting a solid surface by its energy per surface area, which contains three terms, the liquid free surface tension $\gamma_{cm}$, the liquid-solid interfacial tension $\gamma_{cs}$ and a term that depends on the thickness $H$ of the film, $P(H)$. We assume that when the surface of the solid is perfectly flat, defined by the plane $Z=0$, the film has an optimal thickness $H_0$. When the solid surface is rough, with a roughness $h_0(X,Y)$, the position $\zeta(X,Y)$ of the liquid free surface is governed by the minimization of the energy:
\begin{equation}
\int\int{dX dY \left[P(\zeta-h_0)+\frac{1}{2} \gamma_{cm}(\vec{\nabla}\zeta)^2\right]}
\label{E_surf}
\end{equation}
We made the assumption that the curvature of the free surface remains small. For small deformations, $P$ is expanded up to second order around $\zeta-h_0=H_0$. The integral of the zero-order term is a constant, that of the first-order term is equal to zero assuming that the total volume of liquid is constant. Finally the minimization of the energy yields:
\begin{equation}
\frac{1}{\xi_e^2}(\zeta-h_0-H_0)-\vec{\nabla}^2 \zeta = 0
\label{rough}
\end{equation}
where the healing length $\xi_e$ is defined by Eq. \ref{xi_e}. 

When the substrate shows a sinusoidal profile with wavelength $\lambda$ and amplitude $A$, as defined by Eq. \ref{sinus_profile}, Eq. \ref{rough} yields to $\zeta$ also varying sinusoidally, with amplitude $a$ displayed in Eq. \ref{a_over_A_vs_lambda}. Since $a<A$, the thickness of the liquid film is maximum in the valleys, where $\cos(2\pi X/\lambda)=-1$, than on the ridges, where $\cos(2\pi X/\lambda)=+1$, and the difference in the thickness of the film between the valleys and the ridges is equal to:
\begin{equation}
\Delta H =A-a=\frac{A}{1+\left(\frac{\lambda}{2 \pi \xi_e} \right)^2}
\label{eq_delta_H_vs_lambda}
\end{equation}

\subsection*{B. Resolution of the elastic model}

Following \cite{dervaux2015,dervaux20}, we solve the most general linear elastic model described in the main text ( Eqs. \ref{Navier}-\ref{normal_surface_stress} together with \ref{Laplace} and \ref{shear}) by writing the balance of linear momentum, incompressibility condition and boundary conditions in Fourier space. Introducing the Fourier transform $\hat{f}(k,Z)$ of a function $f(X,Z)$  as:

\begin{equation}
\hat{f}(k,Z)=\frac{1}{\sqrt{2 \pi}} \int^{\infty}_{\infty} f(X,Z)e^{ikX}\mbox{d}k
\end{equation}

Using this definition the unknown fields $u$ and $p$ can be expressed in term of the vertical component of the displacement field. The incompressibility condition first yields:

\begin{equation}
\hat{u}(k,Z)=\frac{i }{k} \frac{\partial \hat{v}}{\partial Z}
\end{equation}

\noindent while the X-component of the equilibrium equation implies:

\begin{equation}
\hat{p}(k,Z)=\mu\left( \frac{1}{k^2} \frac{\partial^3 \hat{v}}{\partial Z^3} -  \frac{\partial \hat{v}}{\partial Z}\right)
\end{equation}

\noindent where $\hat{u}$, $\hat{v}$ and $\hat{p}$ are respectively the Fourier transforms of $u$, $v$ and $p$. Using these results, the Y-component of the balance of linear momentum yields a fourth-order ordinary differential equation of $\hat{v}$:

\begin{equation}
\frac{\partial^4 \hat{v}}{\partial Z^4}  - 2k^2 \frac{\partial^2 \hat{v}}{\partial Z^2} + k^4\hat{v} = 0
\end{equation}

This equation is supplemented by four boundary conditions. At $Z=0$ we have:

\begin{eqnarray}
&& \frac{\partial \hat{v}}{\partial Z}=0\\ 
&& \hat{v}(k,0)=\frac{A\sqrt{\pi}}{2\sqrt{2}}\left\{\delta\left(k-\frac{2\pi}{\lambda}\right)+\delta\left(k+\frac{2\pi}{\lambda}\right)\right\}
\end{eqnarray}

\noindent while at the free surface $Z=H$ we have:

\begin{eqnarray}
&&\mu\frac{\partial^2 \hat{v}}{\partial Z^2}+k^2\left(\mu + \sigma_{act} \right) \hat{v}=0\\ 
&&3 k^2 \mu\frac{\partial \hat{v}}{\partial Z}- \mu \frac{\partial^3 \hat{v}}{\partial Z^3}=-\gamma k^4  \hat{v}
\end{eqnarray}

Solving the ordinary differential equation for $\hat{v}$ together with the above four boundary condition, we obtain the following solution, here expressed at the free surface for simplicity:

\begin{eqnarray}
 \hat{v}(k,H)&=\frac{A\sqrt{\pi}}{2\sqrt{2}}\left\{\delta\left(k-\frac{2\pi}{\lambda}\right)+\delta\left(k+\frac{2\pi}{\lambda}\right)\right\} \times \nonumber \\
 & \frac{2(\cosh(Hk)+Hk\sinh(Hk))}{(1+2H^2k^2(1-\frac{\gamma}{\mu H}-\frac{\sigma_{act}}{\mu})+\cosh(2Hk)+\frac{\gamma}{2\mu}\sinh(2Hk)}\nonumber
\end{eqnarray}

\noindent where $\delta$ is the Dirac delta distribution. We can now use Fourrier inversion theorem to find the Z-component of the displacement field in real space (again expressed at the free surface for simplicity): 

\begin{eqnarray}
v(X,H)=&\frac{A}{2}\cos(\frac{2\pi X}{\lambda}) \times \nonumber \\
&  \frac{2(\cosh(Hk)+Hk\sinh(Hk))}{(1+2H^2k^2(1-\frac{\gamma}{\mu H}-\frac{\sigma_{act}}{\mu})+\cosh(2Hk)+\frac{\gamma}{2\mu}\sinh(2Hk)}\nonumber
\end{eqnarray}

Note that this last result can be rewritten under the form presented in equation \ref{zeta1}-\ref{zeta2} or, taking $\sigma_{act}=0$, as \ref{zeta00}-\ref{zeta0}

% BibTeX users please use one of
%\bibliographystyle{spbasic}      % basic style, author-year citations
%\bibliographystyle{spmpsci}      % mathematics and physical sciences
\bibliographystyle{spphys}       % APS-like style for physics
\bibliography{biblio_EPJE_2021}   % name your BibTeX data base

\end{document}